\documentclass[a4paper,10pt,twoside]{cpc-hepnp}

\usepackage{multicol}
\usepackage{graphicx}
\usepackage{booktabs}
\usepackage{amssymb,bm,mathrsfs,bbm,amscd}
\usepackage[tbtags]{amsmath}
\usepackage{lastpage}

\begin{document}

\fancyhead[c]{\small Submitted to Chinese Physics C~~~Vol. XX, No. X (2015)
XXXXXX} \fancyfoot[C]{\small 010201-\thepage}

\footnotetext[0]{Received XX XX 2015}

\title{ Influence of the weakly interacting light U boson on the properties of massive PNS\thanks{Supported by National Natural Science Foundation of China (11175147) }}

\author{%
     HONG Bin£¨ºé±ó£©%
\quad JIA Huan-Yu(¼Ö»ÀÓñ)$^{1)}$\email{hyjia@home.swjtu.edu.cn}%
\quad MU Xue-Ling(ĸѩÁá)
\quad ZHOU Xia(ÖÜϼ)
}
\maketitle

\address{%
Institute for Modern Physics, Southwest Jiaotong University, Chengdu 610031, China\\
}

\begin{abstract}
Considering the octet baryons in relativistic mean field theory and selecting entropy per baryon S=1, we calculate and discuss the influence of U boson on the equation of state, mass-radius, moment of inertia and gravitational redshift of massive protoneutron star (PNS). The effective coupling constant $g_{U}$ of U bosons and nucleons is selected from 0GeV$^{-2}$ to 70GeV$^{-2}$. The results point that U bosons will stiffen the equation of state (EOS). The influence of U bosons on the pressure is more obvious at low density than high density, while, the influence of U bosons on the energy density is more obvious at high density than low density. The U bosons play a significant role in increasing the maximum mass and radius of PNS. When the value of $g_{U}$ changes from 0GeV$^{-2}$ to 70GeV$^{-2}$, the maximum mass of massive PNS increases from 2.11 $M_{\odot}$ to 2.58 $M_{\odot}$, and the radius of PNS corresponding PSR J0348+0432 increases from 13.71 km to 24.35 km. The U bosons will increase the moment of inertia and decrease the gravitational redshift of PNS. For PNS of the massive PSR J0348+0432, the radius and moment of inertia vary directly with $g_{U}$, the gravitational redshift vary inversely with $g_{U}$ approximately.
\end{abstract}



\footnotetext[0]{\hspace*{-3mm}\raisebox{0.3ex}{$\scriptstyle\copyright$}2013
Chinese Physical Society and the Institute of High Energy Physics
of the Chinese Academy of Sciences and the Institute
of Modern Physics of the Chinese Academy of Sciences and IOP Publishing Ltd}%

\begin{multicols}{2}

\section{Introduction}
 The massive neutron stars had been observed. Such as the PSR J1614-2230 was measured to be $1.97\pm0.04$ $M_{\odot}$ using the method of shapiro delay \cite{lab1} and the PSR J0348+0432 to be $2.01\pm0.04$ $M_{\odot}$ with a combination of radio timing and precise spectroscopy of the white dwarf companion by Antoniadis et al \cite{lab2}. Many studies on these massive neutron stars support the stiff equation of state of neutron star matter. For example, Tsuyoshi Miyatsu et al  adopted chiral quark¨Cmeson coupling model within relativistic Hartree¨CFock approximation to reconstruct the EOS for neutron star matter at zero temperature including nuclei in the crust and hyperons in the core, and obtained the resultant maximum mass is 1.95 $M_{\odot}$ which is consistent with the PSR J1614-2230 \cite{lab3}. Xian-Feng Zhao and Huan-Yu Jia attempted to find a possible model in relativistic mean field theory (RMFT) to describe the neutron star of PSR J1614-2230 through adjusting different hyperon coupling parameters \cite{lab4}.

As we know, the cold neutron star is one kind of evolution outcomes of a protoneutron star (PNS) which is formed in the core of a massive star. The properties of the PNS corresponding these massive neutron stars are of practical importance. Ilona Bednarek and Ryszard Manka considered the complete form of the equation of state of strangeness rich PNS to mainly study the influence of the strength of hyperon-hyperon interactions on the properties of the PNS for the neutron star whose mass is below 2$M_{\odot}$ \cite{lab5}. The existence of hyperons will soften the EOS and subsequently decrease the mass of neutron star. So how the hot neutron star matter including more hyperons supports these massive protoneutron stars will be a significant problem, and there is little work to discuss it.

The possible existence of a neutral weakly coupling light spin-1 gauge U boson \cite{lab6}, which is originated from supersymmetric extensions of standard model with an extra U(1) symmetry, has recently attracted much attention due to its multifaceted influences in particle physics, nuclear physics, astrophysics, and cosmology \cite{lab7}. The light U boson, which is proposed beyond the standard model, can play the role in deviating from the inverse square law of the gravity due to the Yukawa-type coupling \cite{lab8,lab9,lab10}. Studying properties of the U boson is thus important for understanding the relevant new physics beyond the standard model. Dong-Rui Zhang et al investigated the effects of the U boson on the nuclear matter equation of state (EOS) and neutron star structure, and showed that the vector U boson can significantly stiffen the nuclear matter EOS, consequently, enhance drastically the maximum mass of neutron stars \cite{lab7}.

There is little work about whether and how a weakly interacting light U boson stiffen the EOS of the hot neutron star and influence the properties of massive PNS. In this paper, we will calculate and discuss the influence of the weakly interacting light U boson on PNS. The paper is organized as follows. In section 2, we give the complete form of relativistic mean field theory (RMFT) at finite entropy including U bosons. In section 3, nucleons and hyperons coupling constants as well as the value of coupling constant of U boson are given. In section 4, some calculation results of U bosons' effect on the massive PNS are given. In section 5, the summary will be presented.

\section{Relativistic mean field theory of hot dense matter and the properties of PNS}

Relativistic mean field theory is an effective field theory of hadron interaction \cite{lab11}. The degrees of freedom relevant to this theory are baryons interacting through the exchange of $\sigma,\omega,\rho$ mesons, of which scalar meson $\sigma$ provides the medium-range attraction, vector meson $\omega$ provides short range repulsive, Vector - isospin vector meson $\rho$ describes the difference between neutron and proton.

We study the properties of the hot neutron star in RMF, so the partition function of system is start point. From the partition function we could get various thermodynamic quantities in equilibrium.

 For grand canonical ensemble, the partition function can be written as:
\begin{equation}
  Z=Tr \{ exp[-(\hat{H}-\mu \hat{N})/T] \},
\end{equation}
where $\hat{H}$ and $\hat{N}$ are the Hamiltonian operator and the particle operator respectively, $\mu$ is the chemical potential, $T$ is the temperature. From the partition function we can get particle population density, the energy density and pressure:
\begin{eqnarray}
   n&=&\frac{T}{V} \frac{\partial lnZ}{\partial \mu},\\
  \varepsilon&=&\frac{T^{2}}{V}\frac{\partial lnZ}{\partial T}+\mu n,
  \\
  P&=&\frac{T}{V} lnZ,
\end{eqnarray}
here, $V$ is the volume.
Considering the baryons $B$ and leptons $l$ as fermions, we can get:
 \begin{eqnarray}\nonumber
lnZ_{B,l}&=&\sum_{B,l} \frac{2J_{B,l}+1}{2\pi^2}\int_{0}^{\infty }k^2dk \{ln[1+e^{-(\varepsilon_{B,l}(k)-\mu_{B,l})/T}]\\
 &+&\frac{V}{T}<\mathcal{L}>,
\end{eqnarray}
where $\varepsilon_{B,l}(k)=\sqrt{k^2+m_{B,l}^2}$ is single particle energy of different momentum $k$ corresponding to different baryon and lepton, $J_{B,l}$ is spin quantum number and $\mu_{B,l}$ is chemical potential of baryon and lepton. $\mathcal{L}$ is the Lagrangian density.

The total partition function $Z_{total}=Z_{B}Z_{l}$, where $Z_{B}$ and $Z_{l}$ are partition function of baryons and the standard noninteracting partition function of leptons respectively. The additional condition of charge neutrality equilibrium is listed as following:
\begin{eqnarray}
\nonumber
  &\sum_{B,l}&\frac{2J_{B,l}+1}{2\pi^{2}}q_{B,l}\int^{\infty}_{0}k^{2}n_{B,l}(k)dk=0,
\end{eqnarray}
where the $n_{B}(k)$ and $n_{l}(k)$ are Fermi distribution function of baryons and leptons respectively. They are given by
\begin{eqnarray}
  n_{i}(k)=\frac{1}{1+exp[(\varepsilon_{i}(k)-\mu_{i})/T]}(i=B,l).
\end{eqnarray}
When neutrinos are not trapped, the set of equilibrium chemical potential relations required by the general condition are
\begin{eqnarray}
 \mu_{i}=b_{i}\mu_{n}-q_{i}\mu_{e},
\end{eqnarray}
where $b_{i}$ is the baryon number of particle $i$ and $q_{i}$ is its charge.

The properties of neutron star at finite temperature can be described by the entropy per baryon. The total entropy per baryon is calculated using $s=(S_{B}+S_{l})/(Tn_{B}$), where $S_{B,l}=P_{B,l}+\varepsilon_{B,l}-\sum_{B,l}\mu_{i}n_{i}$ and the sum are extended over all the baryons and leptons species \cite{lab12}.

The Lagrangian density of neutron star matter is given by \cite{lab13}:
\end{multicols}
\begin{eqnarray}\nonumber
\mathcal{L}&=&
\sum_{B}\overline{\Psi}_{B}(i\gamma_{\mu}\partial^{\mu}-{m}_{B}+g_{\sigma B}\sigma-g_{\omega B}\gamma_{\mu}\omega^{\mu}
-\frac{1}{2}g_{\rho B}\gamma_{\mu}\tau\cdot\rho^{\mu})\Psi_{B}+\frac{1}{2}\left(\partial_{\mu}\sigma\partial^{\mu}\sigma-m_{\sigma}^{2}\sigma^{2}\right)
\nonumber\\
&&-\frac{1}{4}\omega_{\mu \nu}\omega^{\mu \nu}+\frac{1}{2}m_{\omega}^{2}\omega_{\mu}\omega^{\mu}-\frac{1}{4}\rho_{\mu \nu}\cdot\rho^{\mu \nu}+\frac{1}{2}m_{\rho}^{2}\rho_{\mu}\cdot\rho^\mu
-U(\sigma)+\sum_{l=e,\mu}\overline{\Psi}_{l}\left(i\gamma_{\mu}\partial^{\mu}
-m_{l}\right)\Psi_{l}
,\
\end{eqnarray}
\begin{multicols}{2}
where the sum on $B$ runs over the octet baryons $(n,p,\Lambda,\Sigma^{-},\Sigma^{0},\Sigma^{+},\Xi^{-},\Xi^{0})$, and $\Psi_{B}$ is the baryon field operator. The term $U(\sigma)$ stands for the scalar $\sigma$ self-interaction:
\begin{eqnarray}
  U(\sigma)=\frac{1}{3}g_{2}\sigma^{3}+\frac{1}{4}g_{3}\sigma^{4}.
\end{eqnarray}
The last term of Equ.(8) represents the free lepton Lagrangian. In addition, we add in Lagrangian $\mathcal{L}_{\mathrm{u}}$ for the influence of the U bosons. According to the conventional view, the Yukawa-type correction \cite{lab14} to Newtonian gravity resides at the matter part rather than the geometric part. Thus, following the form of the vector meson, $\mathcal{L}_{\mathrm{u}}$ is written as \cite{lab10,lab15}:
\begin{eqnarray}
 \mathcal{L}_{\mathrm{u}}=-\overline{\Psi}_{B}g_{\mathrm{u}}\gamma_{\mu}\mathrm{u}^{\mu}\Psi_{B}-\frac{1}{4}U_{\mu\nu}U^{\mu\nu}+\frac{1}{2}m^{2}_{\mathrm{u}}\mathrm{u}_{\mu}\mathrm{u}^{\mu},
\end{eqnarray}
with $\mathrm{u}$ is the field of U boson, $g_{\mathrm{u}}$ is the coupling constant of U bosons and baryons. $U_{\mu\nu}$ is the strength tensor of the U boson, $m_{\mathrm{u}}$ is the mass of U boson.

The relativistic mean field theory gives the formula of energy density and pressure of a neutron star of finite temperature as follows:
\end{multicols}
\begin{eqnarray}\nonumber
\mathbf{\varepsilon}_{0}&=&\frac{1}{3}g_{2}\sigma^{3}+\frac{1}{4}g_{3}\sigma^{4}+\frac{1}{2}m_{\sigma}^{2}\sigma^{2}+\frac{1}{2}m_{\omega}^{2}\omega_{0}^{2}+\frac{1}{2}m_{\rho}^{2}\rho_{03}^{2}\\
\nonumber
&+&\sum_{B}\frac{2J_{B}+1}{2\pi^{2}}\int_{0}^{\infty}\sqrt{k^{2}+(m^{\ast})^{2}}(exp[(\varepsilon_{B}(k)-\mu_{B})/T]+1)^{-1}k^{2}dk\\
&+&\sum_{l}\frac{2J_{l}+1}{2\pi^{2}}\int_{0}^{\infty}\sqrt{k^{2}+m_{l}^{2}}(exp[(\varepsilon_{l}(k)-\mu_{l})/T]+1)^{-1}k^{2}dk,\\
\nonumber
P_{0}&=&-\frac{1}{3}g_{2}\sigma^{3}-\frac{1}{4}g_{3}\sigma^{4}-\frac{1}{2}m_{\sigma}^{2}\sigma^{2}+\frac{1}{2}m_{\omega}^{2}\omega_{0}^{2}+\frac{1}{2}m_{\rho}^{2}\rho_{03}^{2}\\
\nonumber
&+&\frac{1}{3}\sum_{B}\frac{2J_{B}+1}{2\pi^{2}}\int_{0}^{\infty}\frac{k^{2}}{\sqrt{k^{2}+(m^{\ast})^{2}}}\left(exp[(\varepsilon_{B}(k)-\mu_{B})/T]+1\right)^{-1}k^{2}dk\\
&+&\frac{1}{3}\sum_{l}\frac{2J_{l}+1}{2\pi^{2}}\int_{0}^{\infty}\frac{k^{2}}{\sqrt{k^{2}+m_{l}^{2}}}(exp[(\varepsilon_{l}(k)-\mu_{l})/T]+1)^{-1}k^{2}dk,
\end{eqnarray}
\begin{multicols}{2}
where, $m^{*}=m_{B}-g_{\sigma B}\sigma$ is the effective mass of baryon. In addition, we consider the weakly interaction light vector U bosons in RMFT, the energy density and pressure can be expressed in a simple form \cite{lab16}:
\begin{eqnarray}
  \varepsilon_{UB}=P_{UB}=\frac{1}{2}\frac{g_{\mathrm{u}}^{2}}{m^{2}_{\mathrm{u}}}n_{B}^{2},
\end{eqnarray}
where $n_{B}$ is total number density of baryons. For simplification, we define the effective coupling constants of U boson and baryon as $g_{U}=\frac{g^{2}_{\mathrm{u}}}{m^{2}_{\mathrm{u}}}$. From above forms, the total energy density may be expressed as $\varepsilon=\varepsilon_{0}+\varepsilon_{UB}$ ($\varepsilon_{0}$ denotes the energy density without modified gravitational correction), as well as the pressure $P=P_{0}+P_{UB}$.

Once the equation of state is specified, the mass and radius of neutron star can be obtained by solving the well-known hydrostatic equilibrium equations of Tolman- Oppenheimer-Volkoff(OV) \cite{lab17}.
\begin{eqnarray}
\frac{\mathrm dp}{\mathrm dr}&=&-\frac{\left(p+\varepsilon\right)\left(M+4\pi r^{3}p\right)}{r \left(r-2M \right)},
\\\
M&=&4\pi\int_{0}^{r}\varepsilon r^{2}\mathrm dr
.\
\end{eqnarray}

In a uniformly slow-rotating and axially symmetric neutron star, the moment of inertia is given by the following expression \cite{lab18}:
\begin{eqnarray}
  I\equiv\frac{J}{\Omega}=\frac{8\pi}{3}\int^{R}_{0}r^{4}e^{-\nu(r)}\frac{\bar{\omega}(r)}{\Omega}\frac{(\varepsilon(r)+P(r))}{\sqrt{1-2GM(r)/r}}dr,
\end{eqnarray}
where $J$ is the angular momentum, $\Omega$ is the angular velocity of the stellar, $\nu(r)$ and $\bar{\omega}(r)$ are radially dependent metric functions, $R, M(r), \varepsilon(r)$ and $P(r)$ are the radius, mass, energy density and pressure of stellar. The specific form of $\nu(r)$ is determined by the following expression:
\begin{eqnarray}
\nonumber
  \nu(r)&=&-G\int^{R}_{r}\frac{(M(r)+4\pi x^{3}P(x))}{x^{2}(1-2GM(x)/x)}dx\\
  &+&\frac{1}{2}ln\left(1-\frac{2GM}{R}\right).
 \end{eqnarray}
In particular, the dimensionless relative frequency $\tilde{\omega}(r)\equiv\bar{\omega}(r)/\Omega$ satisfies the following second-order differential equation:
\begin{eqnarray}
  \frac{d}{dr}\left(r^{4}j(r)\frac{d\widetilde{\omega}(r)}{dr}\right)+4r^{3}\frac{dj(r)}{dr}\widetilde{\omega}(r)=0,
\end{eqnarray}
where
\begin{eqnarray}
j(r)=e^{-\nu(r)-\lambda(r)}=
\begin{cases}
  e^{-\nu(r)}\sqrt{1-2GM(r)/r} & r \leqslant R, \\
  1 & r > R.
\end{cases}
\end{eqnarray}
Note that $\widetilde{\omega}(r)$ is subject to the following two boundary conditions:
\begin{eqnarray}
\nonumber
\widetilde{\omega}'(0)=0,\\
\widetilde{\omega}(R)+\frac{R}{3}\widetilde{\omega}'(R)=1.
\end{eqnarray}
With the EOS and the OV equation, Eqs.(16-20) could be solved.

General relativity gives the gravitational redshift of the stellar obeying the relation \cite{lab19}:
\begin{eqnarray}
  z=\left(1-\frac{2GM}{c^{2}R}\right)^{-1/2}-1,
\end{eqnarray}
where the $M, R$ are mass and radius of neutron star.

\section{Coupling constants}
In order to calculate the EOS of PNS matter and its properties, we need three kinds of coupling constants. The first kind is the nucleon coupling constants which could be determined from the saturation properties of nuclear matter. For this study, we choose the nucleon coupling constants to be parameter set GL85 listed in Table~\ref{tab1}, which may well describe the interaction between nucleons \cite{lab13}.
\end{multicols}
\begin{center}
\tabcaption{ \label{tab1}  The coupling constants of the nucleons GL85.}
\footnotesize
\begin{tabular*}{170mm}{@{\extracolsep{\fill}}ccccccc}
\toprule $m$/MeV & $m_{\sigma}$/MeV & $m_{\omega}$/MeV & $m_{\rho}$/MeV & $g_{\sigma}$ & $g_{\omega}$ & $g_{\rho}$ \\
\hline
         939 &500          &782         &770       & 7.9955    &9.1698 &9.7163\\
          $g_{2}$/fm$^{-1}$ &$g_{3}$ & $\rho_{0}$/fm$^{-3}$ & $(B/A)$/MeV & $K$/MeV & $a_{sym}$/MeV & $m^{*}/m$ \\
 \hline
          10.07   &29.262           &0.145         &15.95       & 285    &36.8 &0.77\\
\bottomrule
\end{tabular*}%
\end{center}

\begin{multicols}{2}

The second kind is hyperon coupling constant. For the coupling constants related with hyperons, we define the ratios:
\begin{eqnarray}
x_{\sigma H}&=&\frac{g_{\sigma H}}{g_{\sigma N}},
\\
x_{\omega H}&=&\frac{g_{\omega H}}{g_{\omega N}},
\\
x_{\rho H}&=&\frac{g_{\rho H}}{g_{\rho N}},
\
\end{eqnarray}
where $N$ denotes the nucleons (neutron and proton) and $H$ denotes hyperons ($\Lambda, \Sigma$ and $\Xi$). Hyperon coupling constants can not be decided by saturation properties of nuclear matter, but could be extrapolated through the hypernuclear experimental data. The hypernuclear potential depth in nuclear matter $U^{N}_{H}$, which is known in accordance with available hypernuclear data, serves to strictly correlate the value of $x_{\sigma H}$ and $x_{\omega H}$ \cite{lab20}:
\begin{eqnarray}
  U^{N}_{H}=x_{\omega H}V-x_{\sigma H}S,
\end{eqnarray}
where $S=m-m^{*}$, $V=(g_{\omega}/m_{\omega})^{2}\rho_{0}$ are the values of the scalar and vector field strengths for symmetric nuclear matter at saturation respectively. With $U^{N}_{H}$, if we give the value of $x_{\omega H}$, we can get the value of $x_{\sigma H}$. The experimental data of hypernuclear potential depth of $U^{N}_{\Lambda}, U^{N}_{\Sigma}$ and $U^{N}_{\Xi}$ are \cite{lab21,lab22,lab23,lab24,lab25,lab26}:
\begin{eqnarray}\nonumber
U^{N}_{\Lambda}&=&-30\mathrm{MeV},\\
\nonumber
U^{N}_{\Sigma}&=&+30\mathrm{MeV},\\
\nonumber
U^{N}_{\Xi}&=&-15\mathrm{MeV}.
\end{eqnarray}

In studying the properties of the neutron star with RMF, considerable uncertainty exists in the value of $x_{\omega H}$ \cite{lab27}. In order to calculate the mass of the massive neutron star, such as 2 $M_{\odot}$, we choose $x_{\omega\Lambda}=x_{\omega\Sigma}=x_{\omega\Xi}=1$. This means that we don't consider the difference between hyperons and nucleon coupling with $\omega$.

Then the coupling constants $x_{\sigma\Lambda}, x_{\sigma\Sigma}$ and $x_{\sigma\Xi}$ can be calculated by formulas (25):
\begin{eqnarray}
   x_{\sigma\Lambda}=0.85, x_{\sigma\Sigma}=0.57, x_{\sigma\Xi}=0.78.
   \nonumber
 \end{eqnarray}

 The hyperon coupling constants $x_{\rho\Lambda}, x_{\rho\Sigma}$ and $x_{\sigma\Xi}$ are determined by using SU(6) symmetry \cite{lab28}:
 \begin{eqnarray}
   x_{\rho\Lambda}=0, x_{\rho\Sigma}=2, x_{\sigma\Xi}=1.
   \nonumber
 \end{eqnarray}

Using these coupling constants, we calculate the maximum mass of zero temperature neutron star and the resultant maximum mass is as high as $2.10 M_{\odot}$, which shows that above coupling constants are fit for describing the massive cold neutron stars observed recently. Sequentially, these coupling constants can be extrapolated to study the PNS.

The third kind of coupling constants required is that of U boson coupling with nucleon. Reference \cite{lab10} gives that the effective coupling constant $g_{U}$ is 0-150 GeV$^{-2}$. In this work, we choose the value range of $g_{U}$ to be 0 GeV$^{-2}$-70 GeV$^{-2}$.

\section{Calculation and results}
In this work, we focus on the influence of the strength of effective coupling constants $g_{U}$ on the EOS, mass-radius, moment of inertia and gravitational redshift of massive PNS. Virtually, there are different stages during the PNS evolution, the entropy in the central regions is moderately high and the value of the entropy per baryon is about 1 or 2(in units of Boltzmann's constant), which corresponds to temperatures in the range $T$=20-50 MeV. The first stage corresponds to an entropy per baryon S=1, continuously, the second stage called deleptonization era and corresponds to maximum heat and entropy per baryon S=2 \cite{lab12}. Here, we select the entropy per baryon is of order 1 corresponding the first stage to emphasize the PNS.

\subsection{Equation of state of PNS}
The equation of state of the PNS matter is shown in Fig.~\ref{fig1} and Fig.~\ref{fig2}. In Fig.~\ref{fig1}, we give the influence of different $g_{U}$ on the energy and pressure which as a function of baryon number density. It presents that the pressure and energy density all increase with the baryon number density for different $g_{U}$ under S=1, meanwhile, the higher $g_{U}$ will give higher pressure and higher energy density. In Fig.~\ref{fig2}, it shows that the inclusion of the U boson will stiffen the EOS. This is physically obvious since the vector form of U boson provides an excess repulsion in addition to the vector mesons $\omega$.

The influence of U boson on the EOS can be read from Fig.~\ref{fig1}. In upper panel, at $\rho=0.145$ fm$^{-3}$(saturation density), the value of $g_{U}$ changes from 0 GeV$^{-2}$ to 70 GeV$^{-2}$, the value of $lgP$ increases from 33.84 dyne/cm$^{2}$ to 34.20 dyne/cm$^{2}$, and the increment is 0.36 dyne/cm$^{2}$(1.06\%). At $\rho=0.5$ fm$^{-3}$(around central density), 0 GeV$^{-2}$ to 70 GeV$^{-2}$ of $g_{U}$ gives 35.29 dyne/cm$^{2}$ to 35.43 dyne/cm$^{2}$ of $lgP$, the increment is 0.14 dyne/cm$^{2}$(0.39\%). The effect of $g_{U}$ on the pressure are more obvious at low density than high density. Compared to lower panel, at $\rho=0.145$ fm$^{-3}$, the value of $g_{U}$ changes from 0 GeV$^{-2}$ to 70 GeV$^{-2}$, the value of $lg\epsilon$ increases from 14.40 g/cm$^{-3}$ to 14.41 g/cm$^{-3}$, the increment is 0.01 g/cm$^{-3}$(0.06\%), at $\rho=0.5$ fm$^{-3}$, 0 GeV$^{-2}$ to 70 GeV$^{-2}$ of $g_{U}$ gives 14.98 g/cm$^{-3}$ to 15.03 g/cm$^{-3}$ of $lg\varepsilon$, the increment is 0.05 g/cm$^{-3}$(0.33\%). The effect of $g_{U}$ on the energy density are more obvious at high density than low density.
\begin{center}
\includegraphics[width=8.5cm]{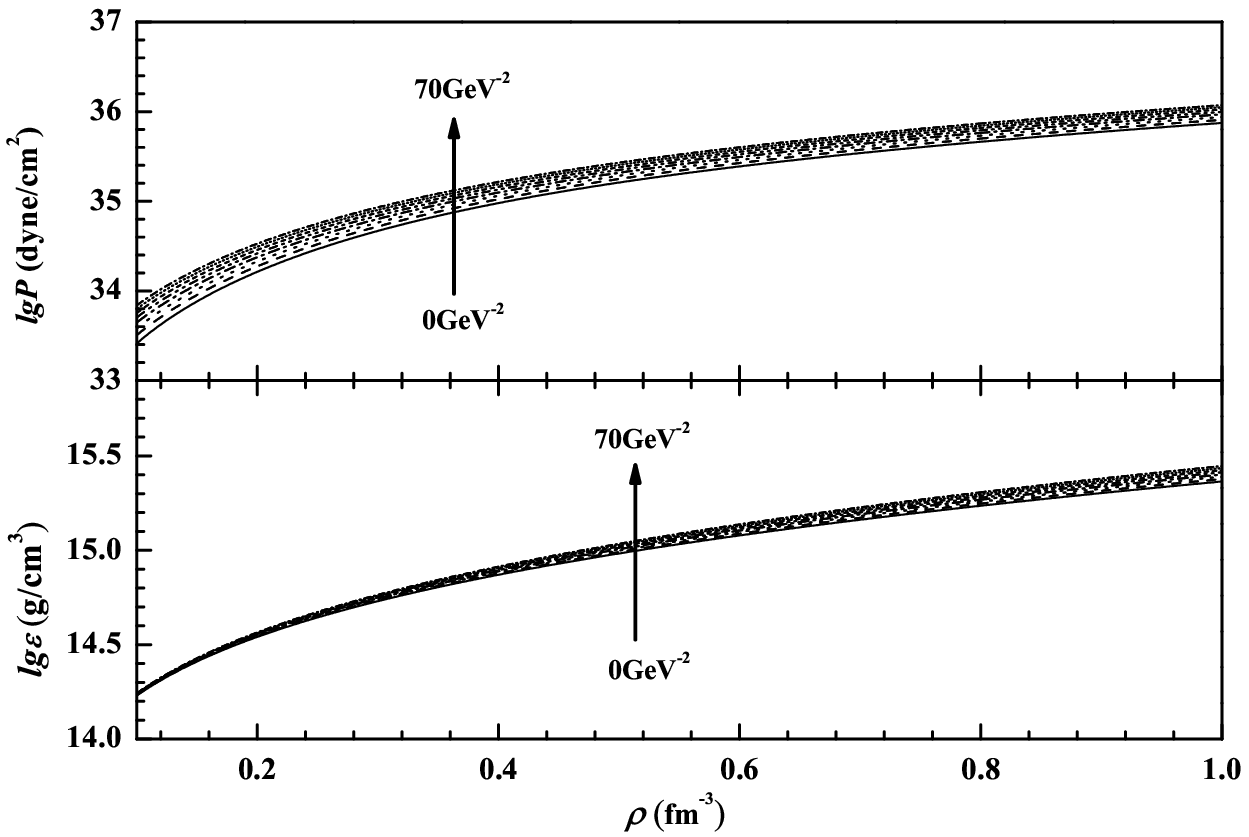}
\figcaption{\label{fig1} Pressure (upper panel) and energy density (lower panel) as a function of baryon number density for different $g_{U}$. }
\end{center}

\begin{center}
\includegraphics[width=8.5cm]{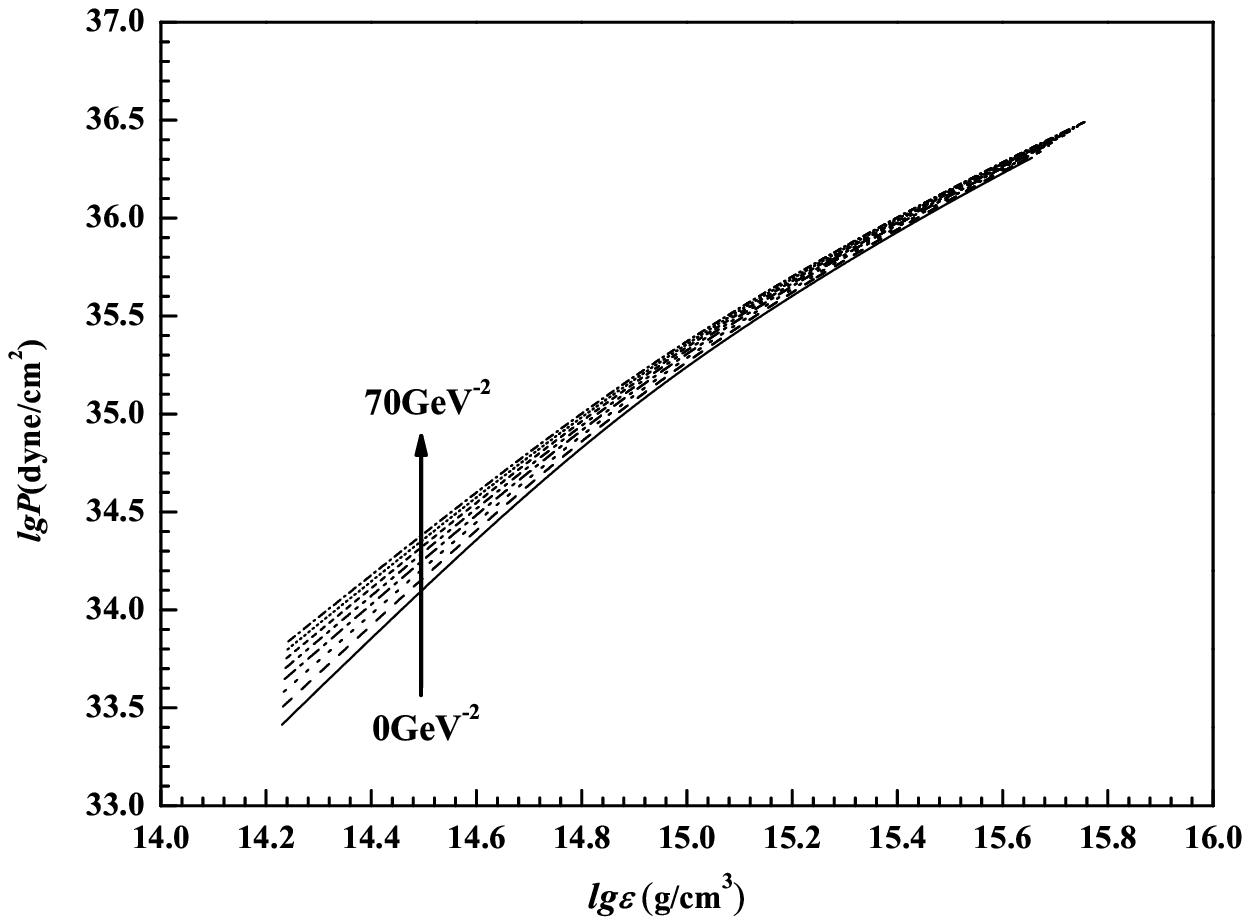}
\figcaption{\label{fig2} The EOS of PNS (S=1) for different $g_{U}$. }
\end{center}

\subsection{Mass-radius of PNS}
Substituting above equation of state into the OV equation, we could solve the masses and radius of the PNS. The resultant masses of PNS are shown in Fig.~\ref{fig3}. In Fig.~\ref{fig3}, the masses of PNS as a function of the central density are given for different effective coupling constant $g_{U}$. The maximum masses of PNS could be read out from Fig.~\ref{fig3}. It is found that the maximum masses of PNS increase significantly with the effective coupling constants. When the value of $g_{U}$ changes from 0 GeV$^{-2}$ to 70 GeV$^{-2}$, the maximum mass of PNS increases from 2.11 $M_{\odot}$ to 2.58 $M_{\odot}$. In these calculations we consider the octet baryons, with each addition of hyperon species, the equation of state is softened for the reason that the fermi pressure of neutrons and protons near the top of their fermi seas is relieved by allowing them to hyperonize to unoccupied low-momentum states \cite{lab13}, consequently leads to decrease the mass. However, the U boson included will stiffen the EOS and increase the mass of neutron star for the reason that it provides extra repulsion in addition to vector meson $\omega$. So this equilibrium result strongly support the existence of the massive PNS whose mass is larger than 2 $M_{\odot}$.

The mass of PNS corresponding to PSR J0348+0432 is marked in Fig.~\ref{fig3}. It is shown that the central density of PNS of PSR J0348+0432 will change with the effective coupling constants. The larger $g_{U}$ is, the lower the central density is. $g_{U}$=0 GeV$^{-2}$, the central density of PNS of PSR J0348+0432 is 0.62 fm$^{-3}$, while $g_{U}$=70 GeV$^{-2}$, the central density becomes 0.21 fm$^{-3}$.

Mass-radius relation is shown in Fig.~\ref{fig4}. In Fig.~\ref{fig4} the masses of PNS as a function of the radius are given with the inclusion of U boson in different effective coupling constants. It can be found that the U bosons will significantly increase the radius of PNS. Here we give the radius of PNS corresponding to PSR J0348+0432. When the value of $g_{U}$ changes from 0 Gev$^{-2}$ to 70 GeV$^{-2}$, the radius increases from 13.71 km to 24.35 km and it is easily seen that if the $g_{U}$ increases to larger than $g_{U}$=70 GeV$^{-2}$, the radius will become bigger than 25 km. It is the reason why we choose the $g_{U}$ at 0-70 GeV$^{-2}$.
\begin{center}
\includegraphics[width=8.5cm]{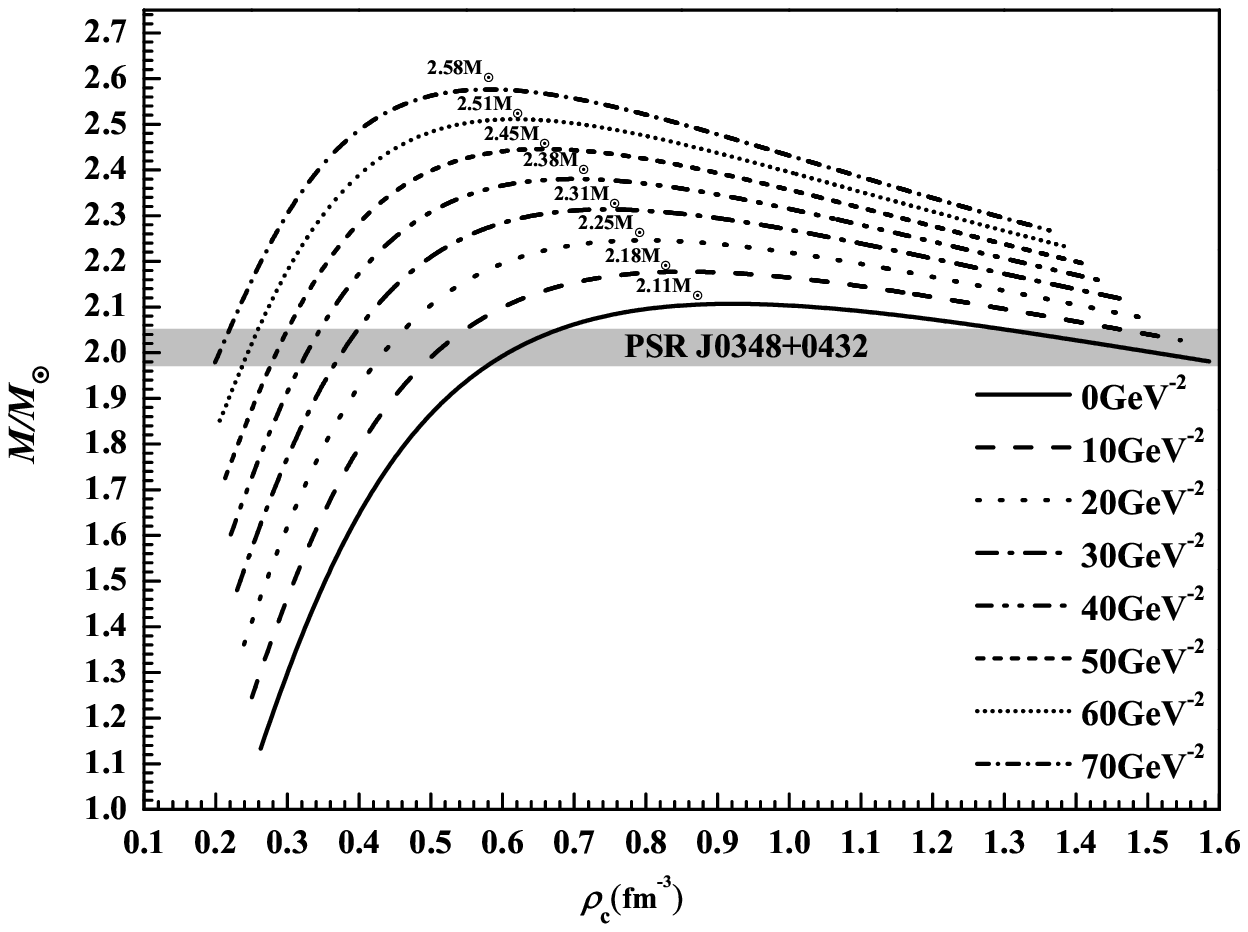}
\figcaption{\label{fig3} The masses of massive PNS for different $g_{U}$. The shaded area corresponds to the mass of PSR J0348+0432. }
\end{center}
\begin{center}
\includegraphics[width=8.5cm]{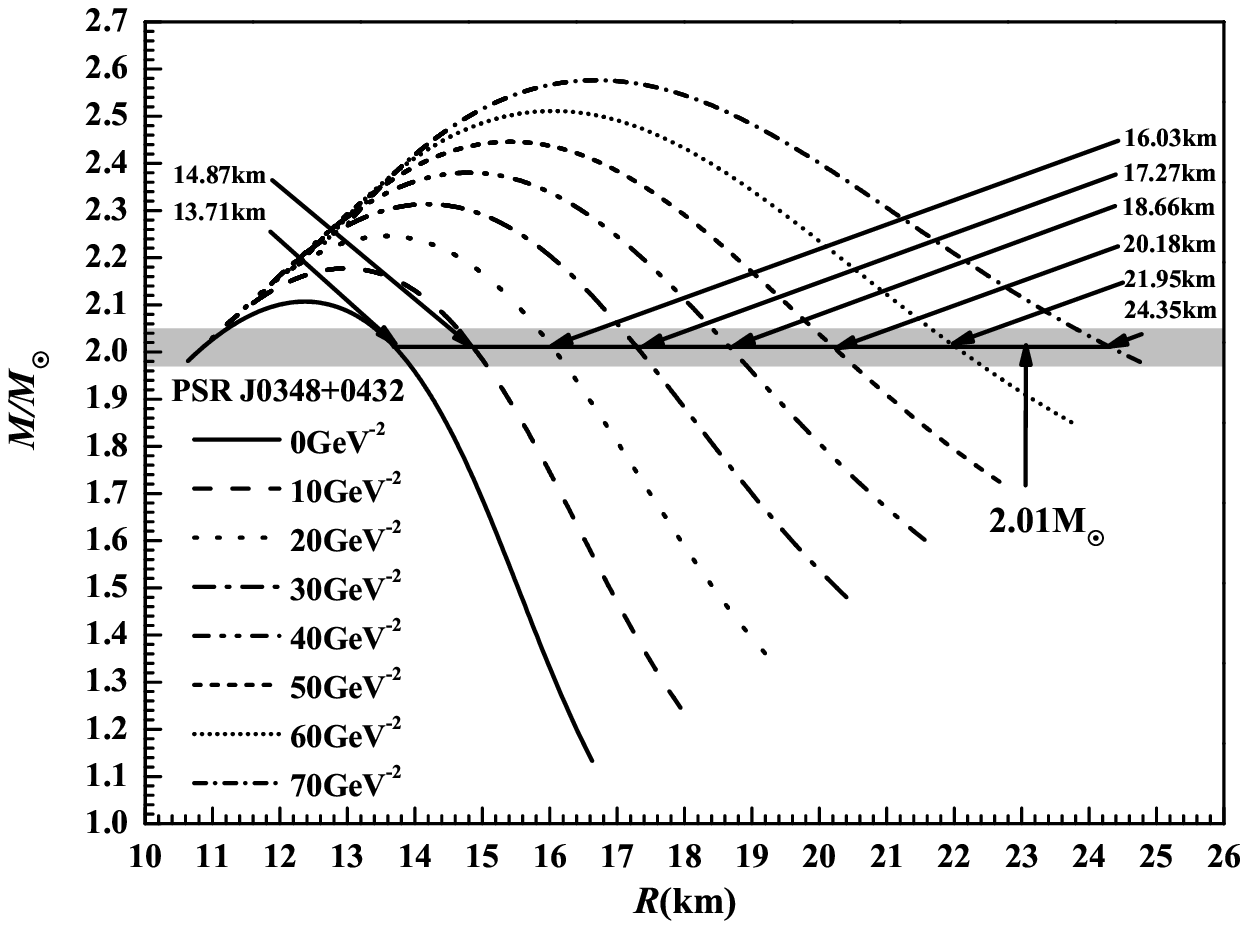}
\figcaption{\label{fig4} The mass-radius relation of massive PNS for different $g_{U}$. The shaded area corresponds to PSR J0348+0432.}
\end{center}
\subsection{The moment of inertia and gravitational redshift of PNS }
The moment of inertia and gravitational redshift are shown in Fig.~\ref{fig5} and Fig.~\ref{fig6}. It can be seen that the moment of inertia increase with the $g_{U}$, but, the gravitational redshift decrease with the $g_{U}$ increases. These due to that the bigger $g_{U}$ will give the larger radius known from the Fig.~\ref{fig4}.
\begin{center}
\includegraphics[width=8.5cm]{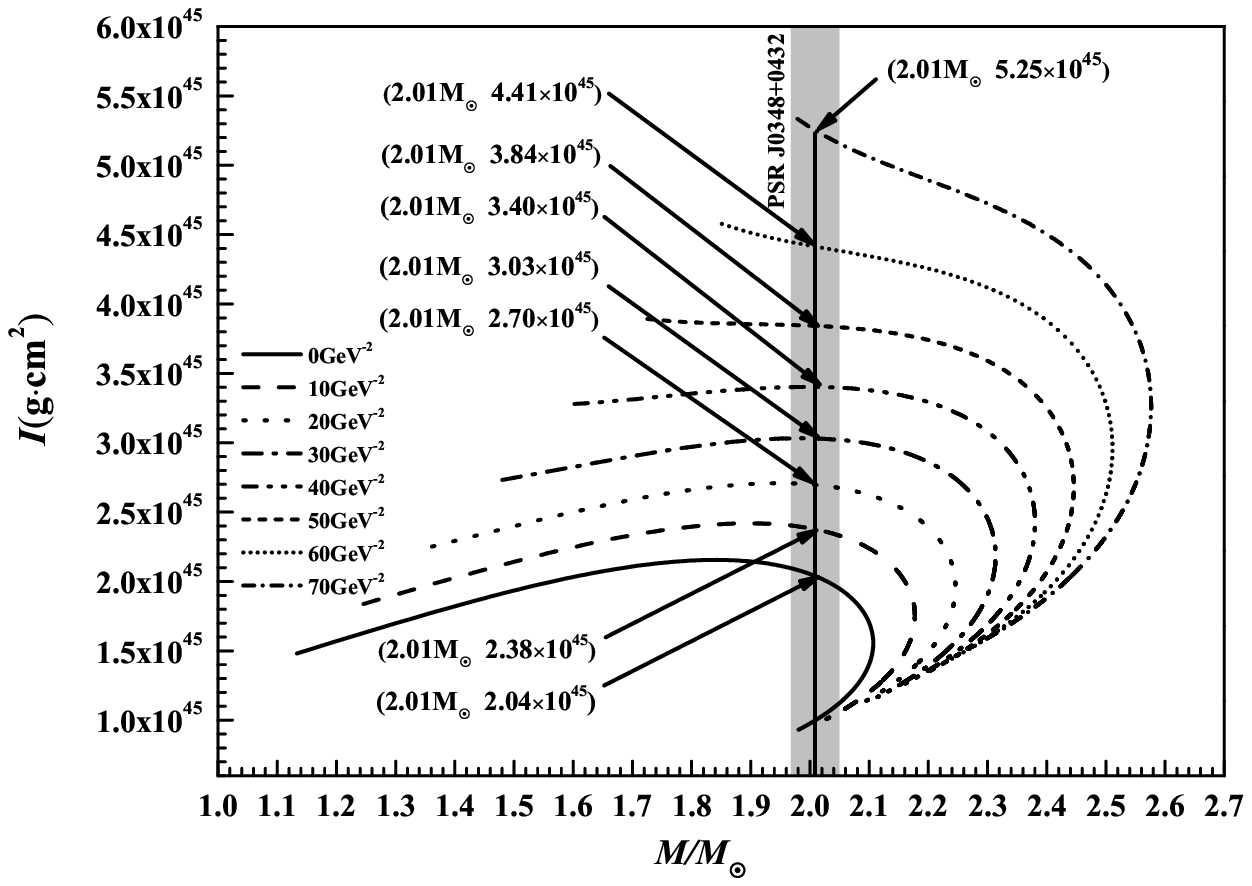}
\figcaption{\label{fig5} The relation between the moment of inertia and mass for different $g_{U}$. The shaded area corresponds to PSR J0348+0432. }
\end{center}
\begin{center}
\includegraphics[width=8.5cm]{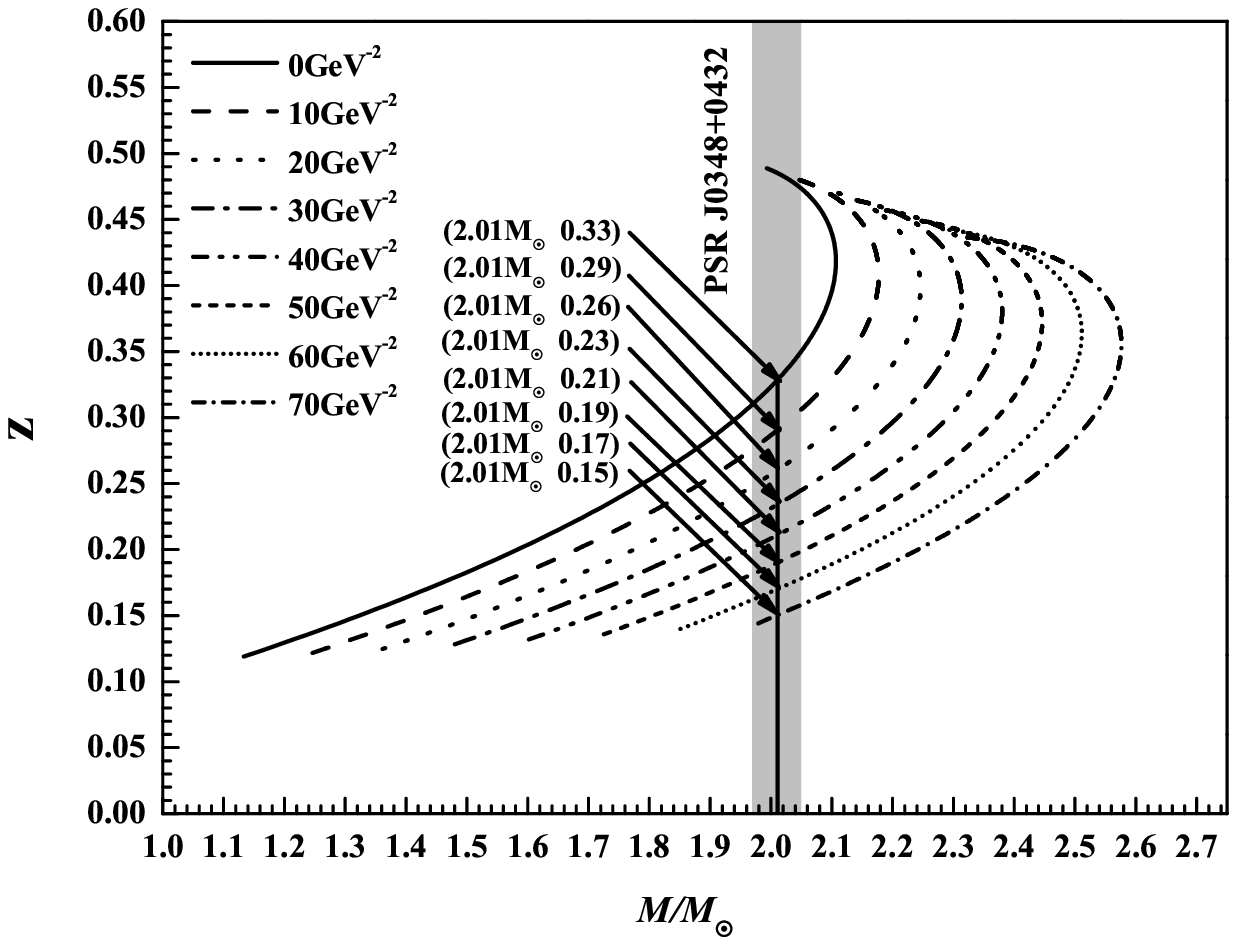}
\figcaption{\label{fig6} The relation between gravitational redshift and mass for different $g_{U}$. The shaded area corresponds to PSR J0348+0432. }
\end{center}

Finally, the change of the properties with $g_{U}$ on massive PNS of PSR J0348+0432 whose mass is 2.01 M$_{\odot}$ is shown in Fig.~\ref{fig7}. The properties include radius, moment of inertia and gravitational redshift. The radius and moment of inertia vary directly with $g_{U}$ , as well as the gravitational redshift vary inversely with $g_{U}$ approximately.
\begin{center}
\includegraphics[width=12cm]{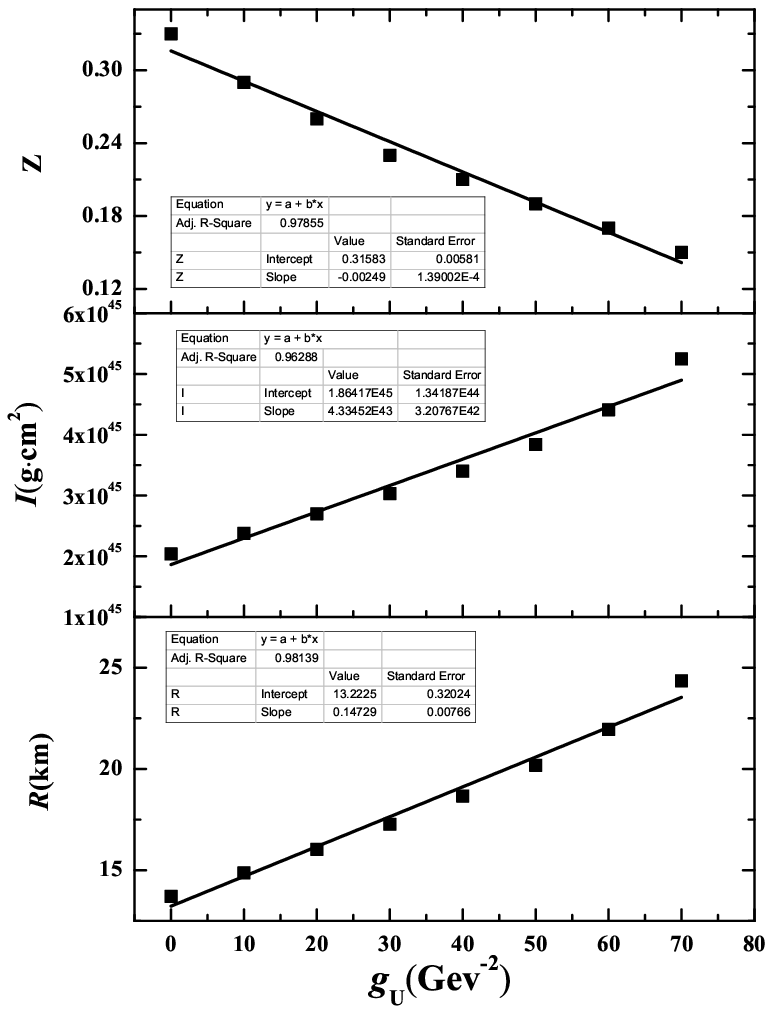}
\figcaption{\label{fig7} The properties of massive PNS of PSR J0348+0432 vs different $g_{U}$.}
\end{center}

\section{Summary}
Basing on relativistic mean field theory, considering the octet baryons and selecting entropy per baryon S=1, we calculate and discuss the influence of U bosons which are weakly coupled to nucleons on massive PNS matter. The effective coupling constant $g_{U}$ of U bosons and nucleons is selected from 0GeV$^{-2}$ to 70GeV$^{-2}$. It is found that the strength of effective coupling constant $g_{U}$ has obvious influence on the EOS of PNS matter, masses, radius, moment of inertia and gravitational redshift of massive PNS. The results point that U bosons will stiffen the EOS, the influence of U bosons on the pressure are more obvious at low density than high density, while, the influence of U bosons on energy density are more obvious at high density than low density. The $g_{U}$ play a significant role in increasing the maximum mass and radius of PNS. When the value of $g_{U}$ changes from 0GeV$^{-2}$ to 70GeV$^{-2}$, the maximum mass of massive PNS increases from 2.11 $M_{\odot}$ to 2.58 $M_{\odot}$, and the radius of PNS corresponding PSR J0348+0432 increases from 13.71 km to 24.35 km. The $g_{U}$ will increase the moment of inertia and decrease the gravitational redshift. For PNS of the massive PSR J0348+0432, the radius and moment of inertia vary directly with $g_{U}$ , the gravitational redshift vary inversely with $g_{U}$ approximately.

\end{multicols}

\vspace{-1mm}
\centerline{\rule{80mm}{0.1pt}}
\vspace{2mm}

\begin{multicols}{2}

\end{multicols}

\clearpage

\end{document}